\documentclass[prl,twocolumn,showpacs]{revtex4}
\usepackage{amsmath}
\usepackage{epsfig}
\begin{document}
\def\b{\bar}
\def\d{\partial}
\def\D{\Delta}
\def\cD{{\cal D}}
\def\cK{{\cal K}}
\def\f{\varphi}
\def\g{\gamma}
\def\G{\Gamma}
\def\l{\lambda}
\def\L{\Lambda}
\def\M{{\Cal M}}
\def\m{\mu}
\def\n{\nu}
\def\p{\psi}
\def\q{\b q}
\def\r{\rho}
\def\t{\tau}
\def\x{\phi}
\def\X{\~\xi}
\def\~{\widetilde}
\def\h{\eta}
\def\bZ{\bar Z}
\def\cY{\bar Y}
\def\bY3{\bar Y_{,3}}
\def\Y3{Y_{,3}}
\def\z{\zeta}
\def\Z{{\b\zeta}}
\def\Y{{\bar Y}}
\def\cZ{{\bar Z}}
\def\`{\dot}
\def\be{\begin{equation}}
\def\ee{\end{equation}}
\def\bea{\begin{eqnarray}}
\def\eea{\end{eqnarray}}
\def\half{\frac{1}{2}}
\def\fn{\footnote}
\def\bh{black hole \ }
\def\cL{{\cal L}}
\def\cH{{\cal H}}
\def\cF{{\cal F}}
\def\cP{{\cal P}}
\def\cM{{\cal M}}
\def\olam{\stackrel{\circ}{\lambda}}
\def\oX{\stackrel{\circ}{X}}
\def\const{{\rm const.\ }}
\def\ik{ik}
\def\mn{{\mu\nu}}
\def\a{\alpha}

\title{Electromagnetic Beams Overpass the Black Hole Horizon}

\author{A Burinskii}
\address{Gravity Research Group NSI Russian Academy of Sciences.
B. Tulskaya 52, Moscow 115191, Russia}

\begin{abstract}
We show that the electromagnetic excitations of the Kerr black
hole have very strong back reaction on metric. In particular, the
electromagnetic excitations aligned with the Kerr congruence form
the light-like beams which overcome horizon, forming the holes in
it, which allows matter to escape interior. So, there is no
information lost inside the black hole. This effect is based
exclusively on the analyticity of the algebraically special
solutions.
\end{abstract}

\pacs{42.50.Lc, 03.70.+k, 11.10.Ef} \maketitle

1. {\bf Introduction.} The problem of the origin of Hawking
radiation \cite{Haw} is not fully understood yet, and many new
suggestions and proposals appears up to now. New arguments on the
role of quantum anomalies in evaporation \cite{RobWil}, and new
derivations of spectrum \cite{Strom,Carl,Carl1,CousHen,Banad}
differ very much from the original Hawking idea and display the
relationships of the black hole physics with the particle physics
and (super)string theory, core of which is based on complex
analyticity of two dimensional conformal field theories with
spherical or $1+1$ topologies,  and on the important role of
diffeomorphism group related with metric deformation.

In particular, in the recent paper by Asthekar, Taveras and
Varadarajan \cite{ATV} the issue of information loss in
two-dimensional space-time is analyzed using the model of
dilatonic gravity model \cite{CGHS}, having  the lagrangian of low
energy string theory. Using a very complicate (but approximate)
analysis authors argue that information is not lost in this model,
due to essential
 peculiarities of null infinity in the quantum space-time with respect
 to the classical one.

In this short note we would like to show that the problem of
information lost is to be resolved already on the classical level
of the Einstein-Maxwell field theory based on the analytical
properties of the algebraically special solutions, in particular,
the Kerr space-time.\fn{The analytic structure of the Kerr
geometry is four-dimensional and based on a twistor analyticity
claimed by the Kerr theorem \cite{BurTwi,Multiks}, which is close
to the Nair \cite{Nair} and Witten \cite{Wit} concept on the role
of twistor space analyticity in quantum theory.}

Main arguments are very simple and based on the fact that the
black hole horizon, even in the classical Einstein-Maxwell field
theory, is very elastic and compliant with respect to
electromagnetic field. For example, the position of the external
horizon of  the Reisner-Nordstr\"om black hole, $r_+ = m + \sqrt
{m^2 - e^2} ,$ is very sensitive to the value of charge $e ,$ and
for $e^2
>m^2 $ horizon disappear at all. Indeed the position of event horizon is
determined by the local values of electromagnetic field. Horizon
is a null surface $S=const.$ determined by metric $g^\mn$ in
accord with the equation \be g^\mn\nabla_\m \nabla _\n S=0 ,
\label{evhor} \ee and the aforementioned dependence of the horizon
for Reisner-Nordstr\"om black hole from charge really is its
dependence from the ratio of the electromagnetic and gravitational
fields in its neighborhood. For more complicate cases, it is
convenient to consider black hole metric in the Kerr-Schild form

\be g_\mn =\eta_\mn - 2H k_\m k_\n, \label{KS}\ee

which is based on the metric of auxiliary Minkowski space-time
$\eta_\mn .$ Here  $k^\m(x), \ x\in M^4$ is the null vector field
which determines symmetry of space, its polarization, and in
particular, direction of gravitational `dragging`. This vector
field is tangent to the Kerr congruence of the geodesic lines for
some especial family of photons. The structure of Kerr congruence
is shown in Fig.1.

\begin{figure}[ht]
\centerline{\epsfig{figure=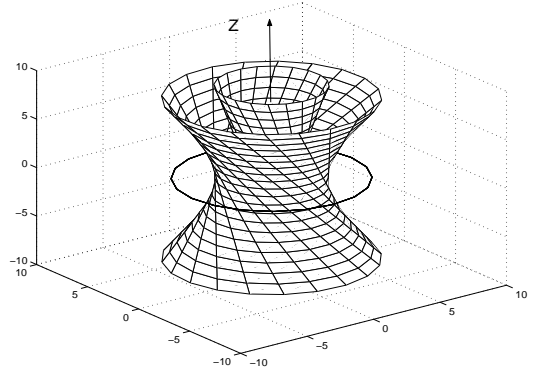,height=5cm,width=7cm}}
\caption{The Kerr singular ring and the Kerr congruence. }
\end{figure}

This congruence is twisting, which determines the complicate form
of the Kerr solution, in spite of the extremely simple form of the
metric (\ref{KS}).

Horizon is determined by function $H ,$ and elasticity of horizon
follows from the form of function $H ,$ which in accord with the
general Kerr-Schild solutions \cite{DKS} is \be H =\frac {mr -
|\psi|^2/2} {r^2+ a^2 \cos^2\theta} , \label{Hpsi} \ee where the
function $\psi\equiv \psi(Y)$ determines electromagnetic field,
and it may be any analytic function of the complex angular
coordinate \be Y=e^{i\phi} \tan \frac {\theta}{2}  \label{Y} ,\ee
which represents a stereographic projection of celestial sphere on
the complex plane.

The Kerr-Newman solution is the simplest solution of the
Kerr-Schild class having $\psi=q=const.,$ where $q$ is the value
of charge. However, any holomorphic function $\psi(Y) $ yields
also an exact solution of this class \cite{DKS}. The form of
horizon for the nontrivial functions $\psi(Y)$ was analyzed in
\cite{BEHM}, and a convincing proof of the elasticity of horizons
with respect to electromagnetic field was obtained. In particular,
it was shown that electromagnetic field corresponding to the
simplest non-trivial functions $\psi(Y)=qY^{\pm 1}$ has a
beam-like form,  which  pierce the horizons, forming a hole
allowing matter to escape interior of black hole. The surfaces of
the event horizons are null ones and obey the differential
equation $(\d_r S)^2 \{ r^2 +a^2 +(q/\tan \frac {\theta} {2})^2
-2Mr \} - (\d_{\theta} S)^2 =0.$ The initially separated
$\pm$-solutions for the internal and external horizons of the
usual black hole turn out to be joined by a tube, conforming a
simply connected surface. Similarly, two boundaries of the
ergosphere, $r_{s+}$ and $r_{s-}$ determined by the equation
$g_{00} =0 $ form a new united surface, being joined by tube which
represents a wormhole to escape interior. The resulting structure
of horizons is illustrated on the Fig.1 (taken from \cite{BEHM}).

\begin{figure}[ht]
\centerline{\epsfig{figure=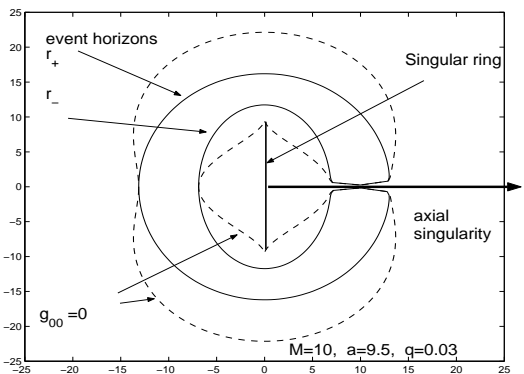,height=5cm,width=7cm}}
\caption{The simplest function $\psi =q/Y$ leads to
electromagnetic beam along z-axis and causes the appearance of
hole in the horizon of rotating black hole.}
\end{figure}

2. {\bf Aligned electromagnetic solutions.} The discussed above
consequence of the simplest beam-like solution may be extended to
the case of arbitrary numbers of beams propagating in different
angular directions $Y_i =e^{i\phi_i}\tan \frac {\theta_i} 2 $

\be \psi (Y) = \sum _i \frac {q_i} {Y-Y_i}. \label{Yi}\ee

In accord with \cite{DKS}, such a function $\psi$ determines
electromagnetic field which describes the corresponding set of
light-like beams along the null rays of the Kerr congruence (see
details in \cite{BEHM}) which have strong back reaction on the
metric, via the function $\psi(Y)$ in (\ref{Hpsi}).
 And again,  for the constant values $q_i $ we will have the exact
 and self-consistent solutions of the full system of Kerr-Schild
 equations, which describe the set of light-like beams destroying
 horizon.

In the recent paper \cite{BEHM1} we showed that this picture is
retained also for the wave  light-like solutions  in the
low-frequency limit.

The main property of the considered beam-like Kerr-Schild
solutions is their analyticity. They are special ones, satisfying
the condition \be k^\m F_{\mn} =0 \label{align} , \ee which means
that they are aligned with the Kerr congruence $k^\m(x)$ depicted
on Fig.1. The appearance of the light-like beam is related with
the complex analyticity of the Kerr-Schild solutions.

Note, that even the very weak aligned excitations destroy horizon,
changing its topology. In particular, considering quantum
evaporation, one has to take into account that the black hole
horizon is very sensitive to vacuum fluctuations, and the related
aligned electromagnetic modes of excitations lead to topological
fluctuations of horizon. It means, that the real horizon subjected
by vacuum fluctuations has to be pierced by a multitude of
migrating holes braking the usual classical image of black hole,
as it is illustrated in Fig.~2.

\begin{figure}[ht]
\centerline{\epsfig{figure=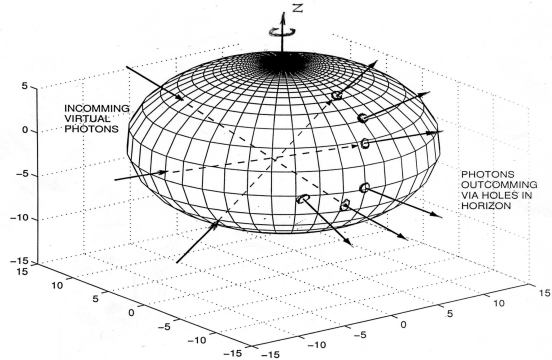,height=6cm,width=8cm}}
\vspace*{-7mm}
\caption{Excitation of a black hole by the
zero-point field of virtual photons forming a set of micro-holes
at its horizon.}
\end{figure}

We are arriving at the conclusion that horizon is not irresistible
obstacle with respect to the beam-like excitations, and
information will not be lost inside black hole. Note, that this
effect is based exclusively on the analyticity of the Kerr-Schild
geometry, caused by algebraically special solutions of the 
Einstein-Maxwell theory.

\end{document}